\documentclass{elsart}

\usepackage{amssymb}
\usepackage{amsmath}
\usepackage{graphicx}

\makeatletter
\def\@date{30 October 2001}
\makeatother

\def\loarrow#1{\raise 1pt\hbox{$\displaystyle\mathop{#1}^{\lower
1pt\hbox{$\scriptstyle\leftarrow$}}$}}
\def\roarrow#1{\raise 1pt\hbox{$\displaystyle\mathop{#1}^{\lower
1pt\hbox{$\scriptstyle\rightarrow$}}$}}

\newcommand{\fgammal}{\roarrow\Gamma_\mathrm{L}}
\newcommand{\fgammar}{\roarrow\Gamma_\mathrm{R}}
\newcommand{\fgammam}{\roarrow\Gamma_\mathrm{M}}

\newcommand{\bgammal}{\loarrow\Gamma_\mathrm{L}}
\newcommand{\bgammar}{\loarrow\Gamma_\mathrm{R}}
\newcommand{\bgammam}{\loarrow\Gamma_\mathrm{M}}

\newcommand{\gammal}{\gamma_\mathrm{L}}
\newcommand{\gammar}{\gamma_\mathrm{R}}
\newcommand{\gammam}{\gamma_\mathrm{M}}

\newcommand{\zetal}{\zeta_\mathrm{L}}
\newcommand{\zetar}{\zeta_\mathrm{R}}

\renewcommand{\etal}{\eta_\mathrm{L}}
\newcommand{\etar}{\eta_\mathrm{R}}
\newcommand{\etam}{\eta_\mathrm{M}}

\newcommand{\ccl}{c_\mathrm{L}}
\newcommand{\ccm}{c_\mathrm{M}}
\newcommand{\ccr}{c_\mathrm{R}}

\renewcommand{\uparrow}{\mathchar"0222}
\renewcommand{\downarrow}{\mathchar"0223}

\begin{document}

\begin{frontmatter}
\title{Incoherent charge transport through molecular wires: interplay of 
Coulomb interaction and wire population}
\author{J\"org Lehmann\corauthref{cor1}},
\corauth[cor1]{Corresponding author. Tel.: +49-821-598-3229; 
fax: +49-821-598-3222}
\author{Gert-Ludwig Ingold}, and
\author{Peter H\"anggi}
\address{Institut f\"ur Physik, Universit\"at Augsburg, 
         Universit\"atsstra{\ss}e~1, D-86135 Augsburg, Germany}

\begin{abstract}
The influence of Coulomb interaction on the electron transport through
molecular wires is studied in the regime of incoherent tunneling. In the
limit of strong Coulomb repulsion, the current for spinless electrons is
determined. It is shown that the voltage profile along the wire crucially
influences the dependence of the current on the wire length. Upon inclusion
of the spin degree of freedom one finds a blocking effect which depends
both on the \emph{interaction strength} and on the \emph{population} 
of the wire. For finite Coulomb interaction, the temperature dependence of 
the blocking is studied and it is shown that several regimes with different 
blocking strength may exist.
\end{abstract}

\end{frontmatter}

\section{Introduction}
\label{sec:intro}

In the last few years, considerable progress towards an electronics based
on single molecules has been made \cite{joach00}. Nanotubes 
\cite{ebbes96,tans97}, fullerenes \cite{joach95,porat97}, and organic molecules
\cite{reed97,datta97,kergu99,reich01} have been contacted and current-voltage 
characteristics were measured. These experimental successes are accompanied
by a vast body of theoretical studies \cite{nitza01} which mostly concentrate 
on coherent transport through the molecules. However, relaxation processes on 
the molecule may lead to \emph{incoherent} transport
\cite{segal00,petro01,petro02} and some of the experimental data may be 
interpreted in this sense \cite{porat97,kergu99}.

Recently, Petrov et al.\ \cite{petro01,petro02} have applied a nonlinear 
quantum kinetic statistical many-body approach in mean-field approximation 
to incoherent transport through short molecular wires in the limit of
infinitely strong Coulomb interaction. Stimulated by this work, we discuss
the charge transport through a single wire in the incoherent regime.
Assuming that relaxation processes on the wire units are much faster than the 
time scale for tunneling between different units, a description in terms
of rate equations for the electronic state of the wire is then adequate.

The main focus of this paper will be on the influence of Coulomb
interaction on the electronic transport. After introducing in
Section~\ref{sec:model} a generic model for the molecular wire and discussing 
the voltage profile across the wire, the limit of strong Coulomb repulsion
will be considered in Section~\ref{sec:nospin}. This allows us to restrict the
calculation to the lowest unoccupied molecular orbitals (LUMO) and to neglect
excitations to higher levels. In particular, this implies that temperature
is sufficiently low so that the Coulomb interaction exceeds the thermal energy.
In Section~\ref{sec:currentws} we will include the spin degree of freedom
for the electrons which had been neglected in Section~\ref{sec:nospin}.
In this context the blocking of spin channels due to Coulomb interaction
will be discussed in detail. Finally, in Section~\ref{sec:finitetemp} we
relax the requirement of strong Coulomb interaction and study the
different regimes arising from the interplay between Coulomb repulsion and
thermal fluctuations.

\section{Model for a molecular wire}
\label{sec:model}

\begin{figure}
  \begin{center}
    \includegraphics[width=7.5truecm]{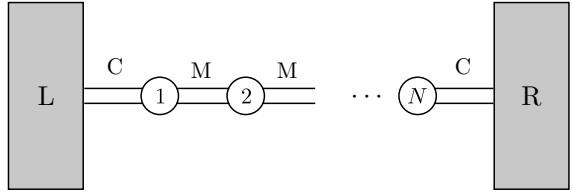}
  \end{center}
  \caption{The molecular wire is modeled by identical wire units represented 
           by circles which are joined by identical molecular bridges ``M''. 
           Two contacts ``C'' provide the connection to the left and right 
           electrodes denoted by ``L'' and ``R'', respectively.}
  \label{fig:wiremodel}
\end{figure}

In this paper, we will focus on generic properties of incoherent 
charge transport through a molecular wire. The model for the wire will 
consequently be kept as simple as possible: a linear array of $N$ identical
wire units numbered consecutively from left to right is joined by $N-1$ 
identical molecular bridges denoted by ``M''. In order to apply a voltage and 
to pass a current through the wire, the setup shown in Fig.~\ref{fig:wiremodel} 
is used where two contacts ``C'' connect the wire to the electrodes marked by
``L'' and ``R''.

Except for Section~\ref{sec:finitetemp}, we will confine the discussion to
the case of strong Coulomb interaction. This prevents the wire from being
charged with more than one electron. Neglecting spin for the moment, the
wire will then always be found in one of $N+1$ states: either the wire is
in its ground state and all wire units are neutral, or exactly one of the
units is charged by an additional electron. We will denote these states by
``0'' for a neutral wire or the number of the unit $i=1,\dots,N$ which 
carries an extra charge.

Since we are interested in the current across the molecule beyond linear
response, we also need to specify the voltage drop along the wire. In
principle, this requires the solution of the coupled Schr\"odinger and
Poisson equations of the electrode-molecule-electrode system \cite{mujic00}, 
which in itself presents a formidable task that we will not try to tackle. 
Instead, we will use the simple model proposed in Refs.~\cite{datta97,tian98}, 
which views the two electrodes as parallel plates forming a capacitor with the 
molecule in between acting like a dielectric. 

\begin{figure}
  \begin{center}
    \includegraphics[width=7.5truecm]{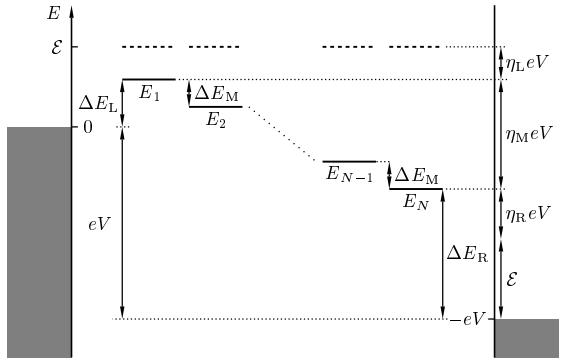}
  \end{center}
  \caption{The energies of the LUMOs of the wire units
  are depicted as dashed lines at energy $\mathcal{E}$ above the
  electrochemical potential in the left electrode. An external voltage 
  $V$ shifts the levels to the positions
  marked by the full lines. The energy differences $\Delta E_\mathrm{L}$,
  $\Delta E_\mathrm{M}$, and $\Delta E_\mathrm{R}$ are defined in
  Eqs.~(\protect\ref{eq:dem})--(\protect\ref{eq:der}). $\eta_\mathrm{L}$,
  $\eta_\mathrm{M}$, and $\eta_\mathrm{R}$ denote the fractions of the total
  applied voltage which drop across the left contact, the wire, and the right
  contact, respectively.}
  \label{fig:voltagedrop}
\end{figure}

Within this picture one obtains a potential profile where the applied 
voltage $V$ drops partially by $\etal V$ and $\etar V$ at the left and
right contact, respectively, and decreases linearly by $\etam V$ along 
the wire. The condition $\etal+\etam+\etar=1$ ensures that we have
accounted for the full voltage drop. This situation is shown in 
Fig.~\ref{fig:voltagedrop} where we have fixed the electrochemical potential
in the left electrode at zero energy. 
The dashed lines at energy $\mathcal{E}$ indicate the position of the LUMOs 
in the absence of an applied voltage. The actual 
levels represented by the full lines at energies
\begin{equation}
  \label{eq:13}
  E_i = \mathcal{E} - \etal e V - \etam \, \frac{i-1}{N-1}\, e V
\end{equation}
are shifted according to the local voltage, thereby displaying the voltage 
profile. 

The current through the wire is governed by the energy differences between
neighboring sites
\begin{equation}
\Delta E_\mathrm{M}=E_i-E_{i+1}=\etam \frac{eV}{N-1}
\label{eq:dem}
\end{equation}
and between the electrochemical potentials in the electrodes and the adjacent
wire units
\begin{align}
\Delta E_\mathrm{L}&=-E_1=\etal e V-\mathcal{E}\label{eq:del}\\
\Delta E_\mathrm{R}&=E_N+eV=\etar eV+\mathcal{E}\ .\label{eq:der}
\end{align}
Particularly at very low temperatures the current depends strongly on the
energy differences (\ref{eq:dem})--(\ref{eq:der}). If one of these energy 
differences is negative, the absence of thermally activated processes will
lead to a suppression of the current. This situation is depicted in
Fig.~\ref{fig:voltagedrop} where $\Delta E_\mathrm{L}<0$. If, on 
the other hand, the energy levels of the
molecular sites lie between the electrochemical potentials in the two 
electrodes, a current may flow even at zero temperature. Apart from
Section~\ref{sec:finitetemp}, we will always have the latter situation in 
mind when discussing zero temperature results. 

\begin{figure}
  \begin{center}
    \includegraphics[width=7.5truecm]{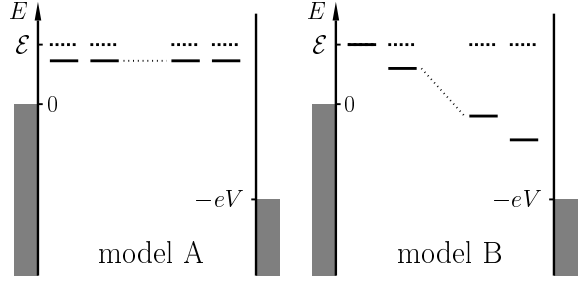}
  \end{center}
  \caption{The positions of the energy levels are shown for models A (left)
   and B (right) of the voltage drop across the system.}
  \label{fig:modelab}
\end{figure}

In the following, the electrochemical potential $-eV$
in the right electrode is assumed to lie lower in energy, thereby
favoring transport from left to right, which we will denote as forward 
direction. Furthermore, we specialize the voltage profile to the two 
limiting cases depicted in Fig.~\ref{fig:modelab}. Within 
\textit{model~A} the voltage is constant along the molecule and drops by equal 
amounts at the two contacts ($\etal=\etar=1/2$, $\etam=0$). In contrast, in 
\textit{model~B} the entire voltage drops linearly along the molecule 
($\etal=\etar=0$, $\etam=1$).

\section{Current for spinless electrons in the presence of strong Coulomb
         interaction}
\label{sec:nospin}

\subsection{Master equation and tunneling rates}
\label{subsec:nospinmaster}

\begin{figure}
  \begin{center}
    \includegraphics[width=5truecm]{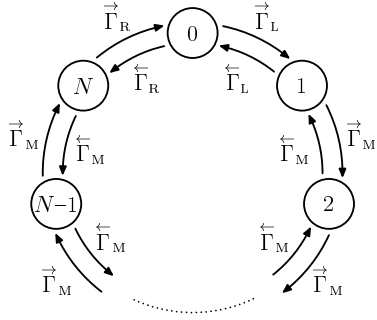}
  \end{center}
  \caption{Transitions between the $N+1$ states of the wire occur either
           by tunneling between adjacent wire units or between the end units
           of the wire and one of the two electrodes.}
  \label{fig:statemodel}
\end{figure}

The incoherent electron transport through an individual wire shall be described 
as sequential tunneling events of an electron between the electrodes and the 
end units of the wire or between adjacent wire units. The dynamics is then
modeled in terms of a Markovian master equation. Fig.~\ref{fig:statemodel}
shows the possible transitions between the $N+1$ different wire states.
The arrows over the corresponding tunneling rates $\Gamma$ indicate the
direction of transport for a geometry depicted in Fig.~\ref{fig:wiremodel},
while the subscripts $L$, $M$, and $R$ correspond to the left contact, a
molecular bridge, and the right contact, respectively.

In order to determine the average current through the wire, we will need
the probabilities $p_i(t)$ to find the system in one of the states $i=0,\dots,
N$. Since these constitute the only possible states, the probabilities should
satisfy the normalization condition
\begin{equation}
  \label{eq:2}
  \sum\limits_{i=0}^N p_i(t) = 1\,.
\end{equation}
The dynamical evolution of the probabilities due to the incoherent
transitions depicted in Fig.~\ref{fig:statemodel} is determined by the
master equation
\begin{align}
  \dot{p}_0(t) & = - (\fgammal + \bgammar ) p_0(t)     + \bgammal p_1(t)
     + \fgammar p_N(t)  \label{eq:1a} \\
  \dot{p}_1(t) & = - (\fgammam + \bgammal ) p_1(t)     + \fgammal p_0(t)
     + \bgammam p_2(t)  \label{eq:1b} \\
  &\;\,\vdots\nonumber\\
  \dot{p}_i(t) & = - ( \fgammam + \bgammam ) p_{i}(t)  + \fgammam p_{i-1}(t)
     + \bgammam p_{i+1}(t) \quad (i=2,\dots,N-1)  \label{eq:1c}\\
  &\;\,\vdots\nonumber\\
  \dot{p}_N(t) & = - ( \fgammar + \bgammam) p_{N}(t)  + \bgammar p_0(t)
  + \fgammam p_{N-1}(t)\,.\label{eq:1d}
\end{align}

From a solution of these equations, the average currents through the left 
and right contacts 
\begin{equation}
I_\mathrm{L}(t) = e(\fgammal p_0 - \bgammal p_1),\quad\quad
I_\mathrm{R}(t) = e(\fgammar p_N - \bgammar p_0),
\label{eq:currcontact}
\end{equation}
and through the molecular bridge between units $i$ and $i+1$
\begin{equation}
I_{\mathrm{M},i}(t) = e(\fgammam p_{i} - \bgammam p_{i+1})
\label{eq:currwire}
\end{equation}
can be obtained. We will be particularly interested in the stationary case
where the probabilities $p_i$ become time-independent. Then, all currents
(\ref{eq:currcontact}) and (\ref{eq:currwire}) will be equal, \emph{i.e.}
\begin{equation}
  \label{eq:curr}
  I=I_\mathrm{L} = I_\mathrm{R} = I_{\mathrm{M},i}\ ,
\end{equation}
thereby ensuring current conservation.

As can be seen from Eqs.~(\ref{eq:1a})--(\ref{eq:1d}), the stationary
probabilities $p_i$ depend only on ratios of transition rates. We therefore 
introduce the ``backward-forward'' ratios
\begin{equation}
  \gammam = \frac{\bgammam}{\fgammam}\ ,\quad
  \gammal = \frac{\bgammal}{\fgammal}\ ,\quad
  \gammar = \frac{\bgammar}{\fgammar}\ ,
  \label{eq:4}
\end{equation}
and the ``branching'' ratios
\begin{equation}
  \zetal = \frac{\fgammam}{\fgammal}\ ,\quad
  \zetar = \frac{\fgammam}{\fgammar}\ .
  \label{eq:5}
\end{equation}
The latter two relate molecular properties to contact properties and therefore
depend on the microscopic details. The ratios (\ref{eq:4}), on the other 
hand, combine backward and forward rates at the same contact or
bridge. In order to ensure thermodynamic equilibrium they are given by
Boltzmann factors at inverse temperature $\beta=1/k_\mathrm{B}T$
\begin{equation}
  \gammam = \e^{-\beta \Delta E_\mathrm{M}}\ ,\ 
  \gammal = \e^{-\beta \Delta E_\mathrm{L}}\ ,\ 
  \gammar = \e^{-\beta \Delta E_\mathrm{R}}\ ,
  \label{eq:5a}
\end{equation}
where the energy differences have been introduced in
Eqs.~(\ref{eq:dem})--(\ref{eq:der}). Since all energy differences have to sum 
up to the difference in electrochemical potential between left and right 
electrode, \emph{i.e.}
\begin{equation}
  \label{eq:5b}
 \Delta
E_\mathrm{L} + (N-1) \Delta E_\mathrm{M} + \Delta E_\mathrm{R} =  eV\ ,
\end{equation}
the ``backward-forward'' ratios obey the relation
\begin{equation}
  \label{eq:5c}
  \gammal \gammam^{N-1} \gammar = \e^{-\beta e V}\ .
\end{equation}

\subsection{Stationary solution}
\label{subsec:nospinstatsol}

In the following, we will restrict ourselves to the stationary case.
Together with the ratios (\ref{eq:4}) and (\ref{eq:5}), the master equation
(\ref{eq:1a})--(\ref{eq:1d}) then turns into
\begin{align}
  - (\zetar + \gammar\zetal) p_0  + \gammal\zetar p_1  + \zetal p_N  & = 0
    \label{eq:3a} \\
  - (\zetal + \gammal) p_1  +  p_0  + \gammam\zetal p_2  & = 0 
    \label{eq:3b} \\
  &\;\,\vdots\nonumber\\
  - ( 1 + \gammam ) p_{i}  +   p_{i-1}   +  \gammam p_{i+1} & = 0\quad 
    (i=2,\dots,N-1)  \label{eq:3c}\\
  &\;\,\vdots\nonumber\\
  - ( 1 + \gammam\zetar) p_{N}  + \gammar p_0   +  \zetar p_{N-1} & = 0\ .
    \label{eq:3d}
\end{align}
Out of these $N+1$ equations only $N$ are linearly independent, so that the
normalization condition~(\ref{eq:2}) is needed to determine the $N+1$ 
stationary probabilities $p_i$. 

It is convenient to start with the ``molecular'' part (\ref{eq:3c}),
which takes on the form of a two-step recursion relation and can thus
readily be solved in terms of the two probabilities $p_1$ and $p_N$ leading
to
\begin{equation}
  \label{eq:7}
  p_i = \frac{1-\gammam^{N-i}}{1-\gammam^{N-1}}\, p_1 - 
        \frac{\gammam^{N-1} - \gammam^{N-i}}{1-\gammam^{N-1}}\, p_N\quad (i=1,\dots,N)\ .
\end{equation}
Inserting this result into Eqs.~(\ref{eq:3b}) and (\ref{eq:3d}) and making
use of the normalization (\ref{eq:2}), one finds for the probability for 
a neutral wire
\begin{equation}
 \label{eq:9}
  p_0 = \left[1 +
              \frac{(\gammam^N-1)(\zetal\gammar + \zetar)+
                    N (\e^{-\beta e V}-1) +
                    \frac{\gammam^N-1}{\gammam-1}(1- \gammal\gammar)
                   }
                   {\gammal (\gammam^{N-1}-1) + (\gammam-1)(\zetal + \zetar\gammal\gammam^{N-1})}
        \right]^{-1}
\end{equation}
while the probability for an electron on site $i$ may be expressed in terms
of $p_0$ as
\begin{equation}
  \label{eq:8}
  p_i = \frac{\e^{-\beta e V} - 1 + 
              \left[(\gammam-1)(\zetal\gammar + \zetar) + 1 - \gammal\gammar
              \right]\gammam^{N-i}
             }
             {\gammal (\gammam^{N-1}-1) + (\gammam-1)(\zetal + \zetar\gammal\gammam^{N-1})} 
        \, p_0\ .
\end{equation}

\subsection{Stationary current}
\label{subsec:nospincurrent}

The stationary solution (\ref{eq:9}) and (\ref{eq:8}) allows to calculate
from any of the expressions (\ref{eq:currcontact}) and (\ref{eq:currwire})
the average current (\ref{eq:curr}). The resulting expression can be cast into 
the form
\begin{equation}
  \label{eq:10}
 I = e\, 
     \frac{1-\e^{-\beta e V}}
          {\frac{\displaystyle \ccl}{\fgammal} + 
           \frac{\displaystyle \ccm}{\fgammam} + 
           \frac{\displaystyle \ccr}{\fgammar}}\,,
\end{equation}
where we have introduced the abbreviations
\begin{align}
  \ccl & = 1 + \frac{\e^{-\beta e V}}{\gammal} \frac{1-1/\gammam^{N}}{1-1/\gammam}\ ,\label{eq:10a}\\
  \ccm & = \gammal \,\frac{1-\gammam^{N-1}}{1-\gammam} + 
            (\gammal\gammar-1)\, \frac{1-\gammam^N}{(1-\gammam)^2} + 
            N \, \frac{1-\e^{-\beta e V}}{1-\gammam}\ , \label{eq:10b}\\
  \ccr & = \frac{\e^{-\beta e V}}{\gammar} + \frac{1-\gammam^N}{1-\gammam}\ .\label{eq:10c}
\end{align}
The current changes its sign when the applied voltage is reversed and at
the same time forward and backward rates as well as left and right contacts
are interchanged. This can be verified by noting that under these changes 
the coefficients (\ref{eq:10a})--(\ref{eq:10c}) are transformed according to
$\ccl/\fgammal\to \e^{\beta e V}\ccr/\fgammar$, 
$\ccr/\fgammar\to \e^{\beta e V}\ccl/\fgammal$, and
$\ccm/\fgammam\to\e^{\beta eV}\ccm/\fgammam$.  

Further insight into the result (\ref{eq:10})--(\ref{eq:10c}) can be
gained by discussing several special cases. An important limit, which will
be particularly relevant in the discussion of the following sections,
is a wire consisting only of one unit and no bridges. Then, the
current simplifies to
\begin{equation}
  \label{eq:12}
  I=
  e\,\frac{\fgammal\fgammar-\bgammal\bgammar}{\fgammal+\bgammal+\fgammar+\bgammar}
\end{equation}
containing forward and backward contributions. For positive energy
differences $\Delta E_\mathrm{L}$ and $\Delta E_\mathrm{R}$, the backward
rates vanish at zero temperature and one is left with the familiar result
\begin{equation}
  \label{eq:12A}
  I= e\,\frac{\fgammal\fgammar}{\fgammal+\fgammar}
\end{equation}
for an sequential incoherent process with rates $\fgammal$ and $\fgammar$.

For more than one wire unit it is instructive to compare the two models for
the voltage profile introduced in Section~\ref{sec:model}. We start by
considering \textit{model~A} where the voltage drops only across the contacts. 
As a consequence, the forward and backward rates on the molecule are equal and 
we have $\gammam=1$. In this case, the coefficients 
(\ref{eq:10a})--(\ref{eq:10c}) become
\begin{align}
  \ccl &= 1 + \frac{e^{-\beta e V}}{\gammal}\, N\ ,\label{eq:12a}\\
  \ccm &= \gammal (N-1) + (1+\e^{-\beta e V})\, \frac{N(N-1)}{2}\ ,\label{eq:12b}\\
  \ccr &= \frac{\e^{-\beta e V}}{\gammar} + N\label{eq:12c}
\end{align}
which is consistent with Eqs.~(\ref{eq:12}) and (\ref{eq:12A}) for $N=1$.
For long wires, the current scales like $1/N^2$ with the number of wire units.
Since forward and backward tunneling rates on the wire are equal, the electron
has equal probability for tunneling in forward and backward direction.
The length dependence of the current can therefore be understood in terms of
a random walk where the average time to diffuse across a wire of length $N$
is proportional to $N^2$.

For \textit{model B}, where the voltage drops entirely along the wire, the
same argument holds provided the thermal energy is much larger than the energy
difference between adjacent sites, $k_\mathrm{B}T\gg \Delta E_\mathrm{M}$.
At low temperatures, however, the situation differs qualitatively since the
backward rates are negligibly small. In order to obtain a finite current
even in the limit of zero temperature, we assume that $\Delta E_\mathrm{L},
\Delta E_\mathrm{R}>0$. At zero temperature, Eqs.~\ref{eq:10})--(\ref{eq:10c})
then yield
\begin{equation}
  \label{eq:11}
  I = e\,\frac{1}{\frac{\displaystyle 1}{\fgammal}+\frac{\displaystyle
  N-1}{\fgammam}+\frac{\displaystyle 1}{\fgammar}}
\end{equation}
which reduces to (\ref{eq:12A}) for $N=1$. For a long wire, the current now
decreases linearly with the number of units or, equivalently, with the length 
of the wire. In this sense, for model B the wire resembles an ohmic resistor. 

The comparison of models~A and B demonstrates that the dependence of the 
current on the wire length at low temperatures may provide qualitative 
information about the voltage drop across the molecule. In any case, however,
incoherent transport leads to an algebraic length dependence of the current
in clear contrast to the exponential dependence characteristic for coherent 
transport \cite{nitza01}.  

\section{Blocking of spin channels by strong Coulomb interaction}
\label{sec:currentws}

So far, we have ignored the spin degree of freedom of the electrons. In this
section, we will take the spin into account but, due to the strong Coulomb
interaction, there will be not more than one electron present on the wire.
Furthermore, we will assume that the two spin directions are equivalent and that
no spin flips occur.

\begin{figure}
  \begin{center}
    \includegraphics[width=5truecm]{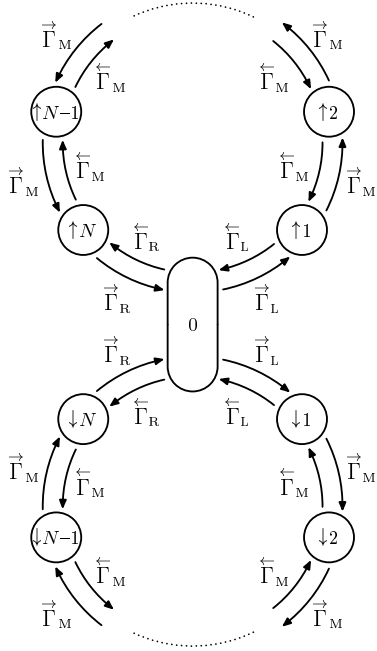}
  \end{center}
  \caption{The situation depicted in Fig.~\protect\ref{fig:statemodel} is
  generalized to include the electron spin. The two rings corresponding to
  the two spin directions are joined via the neutral state ``0'' of the 
  molecule.} 
  \label{fig:statemodelws}
\end{figure}

Inclusion of the electron spin increases the number of possible wire states
to $2N+1$. The wire can now either be neutral (with probability
$p_0(t)$) or there can be an extra electron with spin up or down at site
$i=1,\dots,N$ (with probability $p_{\uparrow,i}(t)$ and
$p_{\downarrow,i}(t)$, respectively). The possible transitions between the
different states are depicted in Fig.~\ref{fig:statemodelws}. We emphasize
that, like in Section~\ref{sec:nospin}, all rates are to be understood as
rates per spin direction. In the presence of a magnetic field, the two
spin directions would no longer be equivalent and the rates would have to
be distinguished by a spin index.

A comparison of Figs.~\ref{fig:statemodel} and \ref{fig:statemodelws} shows that
the two molecular parts are independent of the spin direction and correspond to the
molecular part in the spinless case. The only difference lies in the coupling to
the neutral state ``0''. This motivates to introduce the total probability
$p_i(t) = p_{\uparrow,i}(t)+ p_{\downarrow,i}(t)$ for an electron on site $i$
irrespective of its spin. The resulting master equation including the
normalization condition is of the same form as Eqs.~(\ref{eq:2}) and
(\ref{eq:1a})--(\ref{eq:1d}) in the spinless case. The only difference consists
in a factor of 2 by which the rates $\fgammal$ and $\bgammar$ describing the
charging of the wire have to be multiplied. For these rates, the spin degree
of freedom leads to an additional process allowing to occupy the wire. This is 
not the case for an electron leaving the wire. In this case, the spin is fixed 
and the corresponding rates $\bgammal$ and $\fgammar$ are not doubled. 

The appearance of a factor of 2 may also be understood in terms of the 
density of states in the electrodes which enters the tunneling rates.  
For tunneling of electrons onto the wire the number of initial states is 
increased by the spin degree of freedom thereby leading to an effective 
doubling of the density of states. On the other hand, for tunneling into the 
electrodes the density of final states is not increased since the spin is 
determined by the electron leaving the wire.

Taking into account the factor of 2 in the tunneling rates that populate the 
wire, one finds for the average current
\begin{equation} 
\label{eq:12B} 
I = e (2 \fgammal\, p_0 - \bgammal \, p_1)\ .  
\end{equation}
Here, the probability $p_1$ implicitly contains a factor of 2 since the
electron contributing to the backward current may carry either spin up or
down. Making the required modifications of the rates $\fgammal$ and 
$\bgammar$, the current (\ref{eq:10})--(\ref{eq:10c}) turns into
\begin{equation}
 \label{eq:12C}
 I = e\, 
     \frac{1-\e^{-\beta e V}}
          {\frac{\displaystyle \ccl}{\fgammal} + 
           \frac{\displaystyle \ccm}{\fgammam} + 
           \frac{\displaystyle \ccr}{\fgammar}}
\end{equation}
where the coefficients now are given by
\begin{align}
  \ccl & = \frac{1}{2} + \frac{\e^{-\beta e V}}{\gammal}
  \frac{1-1/\gammam^{N}}{1-1/\gammam}\ ,
  \label{eq:12D}\\
  \ccm & = \frac{\gammal}{2} \frac{1-\gammam^{N-1}}{1-\gammam} + 
            (\gammal\gammar-1)\, \frac{1-\gammam^N}{(1-\gammam)^2} + 
            N \, \frac{1-\e^{-\beta e V}}{1-\gammam}\ ,\label{eq:12E}\\
  \ccr & = \frac{\e^{-\beta e V}}{2\gammar} + \frac{1-\gammam^N}{1-\gammam}\ . \label{eq:12F}
\end{align}
We remark, that due to the mutual blocking of the electrons one does not
obtain an overall factor of 2 in the current relative to the spinless
case (\ref{eq:10})--(\ref{eq:10c}). Instead, the current shows a more
intricate influence of the spin degree of freedom.

\begin{figure}
  \begin{center}
    \includegraphics[width=7.5truecm]{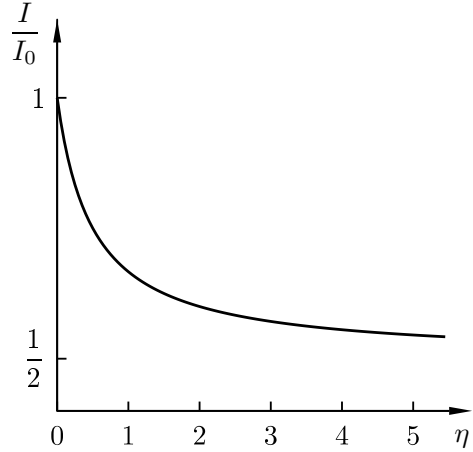}
  \end{center}
  \caption{The current suppression (\ref{eq:blocking}) due to Coulomb
  interaction is plotted as a function of the ratio $\eta$ of the tunneling 
  rates in the left and right contacts.}
  \label{fig:ii0}
\end{figure}

The mechanism of blocking is particularly transparent if the wire consists
of only one site, \emph{i.e.} $N=1$. We restrict ourselves to the case of 
zero temperature where, as in the derivation of Eq.~(\ref{eq:12A}), all 
backward rates vanish. Then the current (\ref{eq:12C}) simplifies to read
\begin{equation}
  \label{eq:12G}
  I = 2 e\, 
     \frac{\fgammal\fgammar}{2 \fgammal + \fgammar}\ .
\end{equation}
This result has to be compared with the case where the two spin directions
lead to independent transport channels so that the expression (\ref{eq:12A})
is multiplied by a factor of 2 yielding the current
\begin{equation}
I_0 = 2e\frac{\fgammal\fgammar}{\fgammal + \fgammar}\ .
\label{eq:i0}
\end{equation}
The effect of blocking due to the Coulomb interaction is quantified by the
ratio of the current (\ref{eq:12G}) in the presence of Coulomb repulsion
and the current (\ref{eq:i0}) in the absence of interaction, \emph{i.e.}
\begin{equation}
\frac{I}{I_0} = \frac{1+\eta}{1+2\eta}\ .
\label{eq:blocking}
\end{equation}
The blocking is completely determined by the ratio 
\begin{equation}
\eta = \frac{\fgammal}{\fgammar}
\end{equation}
of the rates through the left and right contacts
and varies between 1 for $\fgammal\ll\fgammar$ and 1/2 for
$\fgammal\gg\fgammar$ as shown in Fig.~\ref{fig:ii0}. 

This dependence on $\eta$ can be understood in terms of the average population 
of the wire which here is given by $p_1$. From the stationary solution of the 
master equation for the case with spin and $N=1$ one obtains 
\begin{equation}
p_1 = \frac{2\eta}{1+2\eta}\ .
\end{equation}
The blocking effect on the current (\ref{eq:blocking}) may therefore be
directly related to the average population of the wire by
\begin{equation}
\frac{I}{I_0} = 1-\frac{p_1}{2}\ .
\label{eq:ivp1}
\end{equation}
For small $\eta$, or equivalently $\fgammal\ll\fgammar$, the wire is depleted
almost immediately after it has been populated. The average population of the 
wire therefore is very small and the passage of another electron is almost 
never blocked. In the opposite case of large $\eta$, or $\fgammal\gg\fgammar$, 
the right contact represents a bottleneck which leads to a sizable average
population of the wire and therefore to the blocking of transport.

This result clearly shows, that inclusion of the spin degree of freedom
does not necessarily lead to a doubling of the current as would be the case
in the absence of Coulomb interaction on the wire. Apart from the interaction
strength, the average population of the wire plays a decisive role and blocking
is strongest when the right contact increases the population by acting as
a bottleneck.

\section{Temperature dependence of blocking}
\label{sec:finitetemp}

The calculation of the current-voltage characteristics presented so far was
based on the assumption that at any given time there is at most one extra
electron on the wire. Although it is straightforward to generalize the rate
equations (\ref{eq:1a})--(\ref{eq:1d}) to more than one electron by
extending the state space, in general one has to solve the rate equations
numerically. 

\begin{figure}
  \begin{center}
    \includegraphics[width=4.5truecm]{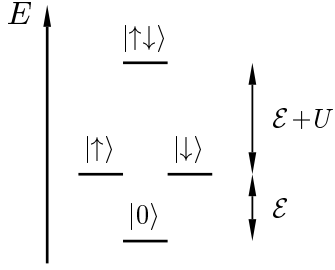}
  \end{center}
  \caption{Many-particle states for one site with $n=0, 1,$ and 2 electrons. 
    $\Delta$ is the one-particle energy of the states $|\uparrow\rangle$ 
    and $|\downarrow\rangle$ and $U$ is the Coulomb interaction energy
    between the electrons occupying the site.}
  \label{fig:twoelec}
\end{figure}

Some insight into the validity of the limit of strong Coulomb interaction
can be gained by allowing a second electron on the wire. Then, an analytical
treatment is still possible if we restrict the length of the wire to one site, 
\emph{i.e.} $N=1$. We will assume that this very site may be occupied
by at most two electrons of opposite spin. Then, there exist
four different wire states, $|0\rangle$, $|\uparrow\rangle$, 
$|\downarrow\rangle$, and $|\uparrow\downarrow\rangle$, which correspond to 
no electron, one electron with spin up or down, and two electrons, 
respectively. The energies of these four states are schematically shown in
Fig.~\ref{fig:twoelec} where we allow for an interaction energy $U$ if two
electrons are present on the wire. By changing both, the temperature as well as 
the one-electron energy $\mathcal{E}$ relative to the left electrochemical 
potential, we may choose the levels which are relevant for the 
transport through the wire.

The discussion in the last section has shown that the effect of blocking is 
most pronounced when the tunneling rates through the right contact are much 
smaller than through the left one. We therefore shall concentrate on this 
limit. Then, the master equation for the probabilities $p_n$ of being in a 
state with $n=0,1,$ or $2$ electrons depends only on the rates through the left 
contact. These tunneling rates may be identified by the change of the number
of electrons. For example, $\Gamma_{2\to 1}$ refers to the backward rate
through the left contact with initially two electrons on the molecule. With
this notation the master equation reads
\begin{align}
  \dot{p_0}(t) & = -2 \Gamma_{0\to 1} \, p_0(t) + \Gamma_{1\to 0} \, p_1(t)\label{eq:a1}\\
  \dot{p_1}(t) & = -(\Gamma_{1\to 0} + \Gamma_{1\to 2}) \,  p_1(t) 
                + 2 \Gamma_{0\to 1} \,p_0(t)  + 2 \Gamma_{2\to 1} \, p_2(t)\label{eq:a2}\\
  \dot{p_2}(t) & = -2 \Gamma_{2\to 1}  \, p_2(t) + \Gamma_{1\to 2} \, p_1(t)\label{eq:a3}
\end{align}
with the normalization condition
\begin{equation}
  p_0(t) + p_1(t) + p_2(t) = 1\ . \label{eq:a4}
\end{equation}
The rates $\Gamma_{0\to 1}$ and $\Gamma_{2\to1}$ are each multiplied by a 
factor of 2, thereby accounting for the two spin directions possible in these 
processes.  The probability $p_1$ of finding one electron of arbitrary spin on 
the wire implicitly contains a factor of 2 in the same sense as was discussed 
after Eq.~(\ref{eq:12B}).

If the probabilities are calculated from the master equation
(\ref{eq:a1})--(\ref{eq:a3}), thereby neglecting tunneling through the right
contact, the current has to be evaluated at that contact. We will assume that 
the electrochemical potential in the right electrode is so low that the 
forward rates $\fgammar$ through the right contact are approximately 
independent of the occupation of the molecular site and that backward rates may 
be neglected. Then, the stationary current is obtained as
\begin{equation}
I = e\fgammar(p_1+2p_2)\ ,
\label{eq:currft}
\end{equation}
where the factor of 2 accounts for the two spin directions available when
one of two electrons is leaving the molecule.

Solving the master equation (\ref{eq:a1})--(\ref{eq:a3}) in the stationary 
limit, we obtain from (\ref{eq:currft}) for the current
\begin{equation}
  \label{eq:a5}
  I= e\fgammar\dfrac{1+\dfrac{\Gamma_{1\to2}}{\Gamma_{2\to1}}}
     {1+\dfrac{\Gamma_{1\to0}}{2\Gamma_{0\to1}}+
        \dfrac{\Gamma_{1\to2}}{2\Gamma_{2\to1}}}\ .
\end{equation}
In order to study the effect of blocking, we have to compare this current
with the current $I_0$ in the absence of Coulomb interaction. In this case
$\Gamma_{1\to2}=\Gamma_{0\to1}$ and $\Gamma_{2\to1}=\Gamma_{1\to0}$ and we
find
\begin{equation}
  I_0= e\fgammar\dfrac{1+\dfrac{\Gamma_{0\to1}}{\Gamma_{1\to0}}}
     {1+\dfrac{\Gamma_{1\to0}}{2\Gamma_{0\to1}}+
        \dfrac{\Gamma_{0\to1}}{2\Gamma_{1\to0}}}\,.
\label{eq:a6}
\end{equation}
This result still depends on temperature and on the energy $\mathcal{E}$ of the 
one-electron level relative to the left electrochemical potential.

The ratio $I/I_0$ depends only on rate ratios which, as in
Section~\ref{subsec:nospinmaster}, may be expressed in terms of Boltzmann
factors. For single occupation of the molecule, we have
\begin{equation}
\dfrac{\Gamma_{0\to1}}{\Gamma_{1\to0}}=\mathrm{e}^{-\beta\mathcal{E}}\ .
\end{equation}
Double occupancy costs the additional interaction energy $U$, so that
\begin{equation}
\dfrac{\Gamma_{1\to2}}{\Gamma_{2\to1}}=\mathrm{e}^{-\beta(\mathcal{E}+U)}\ .
\end{equation}
The blocking due to Coulomb interaction is then described by
\begin{equation}
\dfrac{I}{I_0} = 1 - \dfrac{1-\mathrm{e}^{-\beta U}}
      {\mathrm{e}^{\beta\mathcal{E}}+2+\mathrm{e}^{-\beta(\mathcal{E}+U)}}\ .
\label{eq:block_t}
\end{equation}

\begin{figure}
  \begin{center}
    \includegraphics[width=7.5truecm]{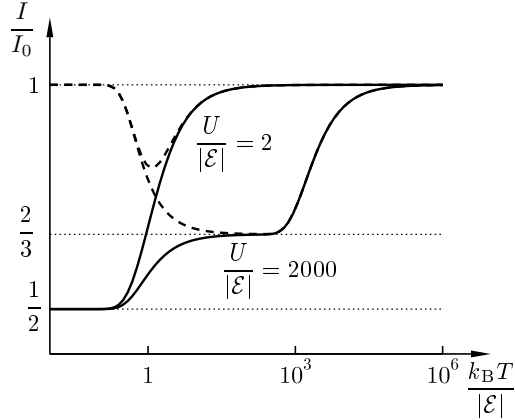}
  \end{center}
  \caption{The current (\protect\ref{eq:a5}) through the wire normalized 
     with respect to the current (\protect\ref{eq:a6}) in the absence of 
     Coulomb interaction is plotted as a function of the temperature $T$ in 
     units of $|\mathcal{E}|$ for $U/|\mathcal{E}|=2$ and 2000. For each
     value of $U/|\mathcal{E}|$, a full line and a dashed line are shown,
     corresponding to $\mathcal{E}<0$ and $\mathcal{E}>0$, respectively.}
  \label{fig:ii0_t}
\end{figure}

Figure~\ref{fig:ii0_t} depicts this result as a function of
$k_\mathrm{B}T/|\mathcal{E}|$ for $U/|\mathcal{E}|=2$ and 2000.
The full lines correspond to the case, where the one-electron energy lies
below the left electrochemical potential, \emph{i.e.} $\mathcal{E}<0$.
The dashed line refers to the opposite case with $\mathcal{E}>0$. As a 
function of temperature, we can distinguish three different regimes.

In the limit of zero temperature, one finds for $\mathcal{E}<0$, and 
$U>-\mathcal{E}$ an interaction induced reduction of the current by a 
factor 1/2. This is the blocking discussed already in 
Section~\ref{sec:currentws}.
For $\mathcal{E}+U<0$, double occupancy is possible even in the zero
temperature limit and no blocking occurs. As can be seen from the dashed
lines in Fig.~\ref{fig:ii0_t}, blocking is also absent for $\mathcal{E}>0$.
In this case, the current as well as the occupation of the molecule 
approach zero exponentially for low temperatures. In analogy to the reasoning
presented in Section~\ref{sec:currentws}, blocking thus becomes ineffective.

For $U\gg|\mathcal{E}|$, a second regime appears at
intermediate temperatures $|\mathcal{E}|\ll k_\mathrm{B}T\ll U$. 
For $U/|\mathcal{E}|=2000$, Fig.~\ref{fig:ii0_t} indeed shows a clear plateau 
at a blocking factor of 2/3. This can be attributed to the fact that for
$k_\mathcal{B}T\gg |\mathcal{E}|$ the backward rate $\Gamma_{1\to0}$ is of 
the order of the forward rate $\Gamma_{0\to1}$. This reduces the population 
of the molecule and therefore blocking is less effective than at lower
temperatures.

The reduction by a factor 2/3 can be understood even without recourse to 
Eq.~(\ref{eq:block_t}). For equal forward and backward rates, the
probabilities for an empty wire or a wire occupied with one electron of given 
spin are equal, leading to probabilities $p_0=1/3$, $p_1=2/3$ and $p_2=0$. 
From Eq.~(\ref{eq:currft}) one therefore finds $I=(2/3)e\fgammar$.
In the absence of Coulomb interaction, double occupancy is allowed and one
finds all four possible states with equal probability leading to $p_0=1/4$,
$p_1=1/2$ and $p_2=1/4$. This yields the current $I_0=e\fgammar$, and thus the 
blocking factor reads $I/I_0=2/3$.

Finally, in the high-temperature regime, where 
$k_\mathrm{B}T\gg |\mathcal{E}|, U$, the thermal energy is much bigger than 
the interaction energy $U$ so that the latter becomes irrelevant. Therefore,
the Coulomb interaction can no longer lead to blocking of the electronic 
transport.

\section{Conclusions}
\advance\tolerance by 400
The influence of Coulomb interaction on incoherent electronic transport 
through a molecular wire has been found to be multifaceted. In the regime
of strong Coulomb interaction, the current of spinless electrons shows
an algebraic dependence on the wire length characteristic for incoherent
transport. It was found, however, that this length dependence varies as
a function of the voltage profile across the system. At low temperatures
a $1/N^2$ behavior indicates the absence of a voltage drop along the molecule
while in the presence of a voltage drop a $1/N$ behavior is expected.

\advance\tolerance by -400
Inclusion of the spin degree of freedom in the absence of Coulomb interaction
leads to an increase of the current by a factor of 2. Coulomb interaction,
on the other hand, should lead to a reduction of the current. It was found
that this blocking effect not only depends on the interaction strength but
also on the average population of the wire and thus on the tunneling rates
(cf.\ Fig.~\ref{fig:ii0}).

For finite Coulomb interaction, the amount of blocking also depends on
temperature. Between the limits of suppression of the current by a factor
of 2 at zero temperature and the absence of blocking at high temperatures
an intermediate regime may exist. There, Coulomb interaction prohibits
double occupancy. Nevertheless the current is only reduced by a factor of
2/3 (cf.\ Fig.~\ref{fig:ii0_t}). This is a consequence of the competition 
between backward and forward rates, which equally populates the two 
one-electron levels. 

As a final remark, we mention that even though we have restricted ourselves
to electronic transport, the approach presented here is also applicable to
hole transport. Depending on the situation, holes may be the dominant charge 
carriers; the results presented above may easily be adapted to this case. 
On the other hand, if both, holes and electrons, contribute to the transport, 
an extension similar to the one presented in Section~\ref{sec:currentws}, 
where spin is included, becomes necessary. 

\ack
\advance\tolerance by 400
The authors would like to acknowledge useful discussions with I. Goychuk.
This work has been supported by the Deutsche Forschungsgemeinschaft
through Sonderforschungsbereich 486, project A10.

\advance\tolerance by -400


\begin{thebibliography}{00}
\bibitem{joach00}
C.~Joachim, J.K.~Gimzewski, A.~Aviram, Nature 408 (2000) 541.

\bibitem{ebbes96}
T.W.~Ebbesen, H.J.~Lezec, H.~Hiura, J.W.~Bennett, H.F.~Ghaemi, T.~Thio,
Nature 382 (1996) 54.

\bibitem{tans97}
S.J.~Tans, M.H.~Devoret, H.J.~Dai, A.~Thess, R.E.~Smalley, L.J.~Geerligs,
C.~Dekker, Nature 386 (1997) 474.

\bibitem{joach95}
C.~Joachim, J.K.~Gimzewski, R.R.~Schlittler, C.~Chavy, Phys.\ Rev.\ Lett.\
74 (1995) 2102.

\bibitem{porat97}
D.~Porath, Y.~Levi, M.~Tarabiah, O.~Millo, Phys.\ Rev.\ B 56 (1997) 9829.

\bibitem{reed97}
M.A.~Reed, C.~Zhou, C.J.~Muller, T.P.~Burgin, J.M.~Tour, Science 278 (1997)
252.

\bibitem{datta97} 
S.~Datta, W.~Tian, S.~Hong, R.~Reifenberger, J.I.~Henderson, C.P.~Kubiak,
Phys.\ Rev.\ Lett.\ 79 (1997) 2530.

\bibitem{kergu99}
C.~Kergueris, J.-P.~Bourgoin, S.~Palacin, D.~Esteve, C.~Urbina, M.~Magoga,
C.~Joachim, Phys.\ Rev.\ B 59 (1999) 12505.

\bibitem{reich01}
J.~Reichert, R.~Ochs, D.~Beckmann, H.B.~Weber, M.~Mayor,
H.~v.~L{\"o}hneysen, cond-mat/0106219.

\bibitem{nitza01}
for a review see e.g. A.~Nitzan, Ann.\ Rev.\ Phys.\ Chem.\ 52 (2001) 681.

\bibitem{segal00}
D.~Segal, A.~Nitzan, W.B.~Davis, M.R.~Wasielewski, M.A.~Ratner,
J. Phys.\ Chem.\ B 104 (2000) 3817.

\bibitem{petro01}
E.G.~Petrov, P.~H{\"a}nggi, Phys.\ Rev.\ Lett.\ 86 (2001) 2862.

\bibitem{petro02}
E.G.~Petrov, V.~May, P.~H{\"a}nggi, this issue.

\bibitem{mujic00}
V.~Mujica, A.E.~Roitberg, M.~Ratner, J. Chem.\ Phys.\ 112 (2000) 6834.

\bibitem{tian98}
W.~Tian, S.~Datta, S.~Hong, R.~Reifenberger, J.I.~Henderson, C.P.~Kubiak,
J. Chem.\ Phys.\ 109 (1998) 2874.
 
\end{thebibliography}
\end{document}